\documentclass{article}
\usepackage{amssymb}
\usepackage[dvips]{graphics,epsfig}

\newcommand{\textin}[1]{\mbox{\scriptsize{#1}}}
\newcommand{\case}[2]{{\textstyle \frac{\mathstrut #1}{\mathstrut #2}}}
\newcommand{\bbox}[1]{\mbox{\boldmath $#1$}}

\begin{document}

\title{Monte Carlo simulation of granular fluids}
\author{Jos\'e Mar\'{\i}a Montanero\\
Departamento de Electr\'onica e Ingenier\'{\i}a Electromec\'anica,\\
Universidad de Extremadura, E-06071 Badajoz, Spain}

\date{\today}
\maketitle

\begin{abstract}
An overview of recent work on Monte Carlo simulations of a granular binary mixture is presented. The results are obtained numerically solving the Enskog equation for inelastic hard-spheres by means of an extension of the well-known direct Monte Carlo simulation (DSMC) method. The homogeneous cooling state and the stationary state reached using the Gaussian thermostat are considered. The temperature ratio, the fourth velocity moments and the velocity distribution functions are obtained for both cases. The shear viscosity characterizing the momentum transport in the thermostatted case is calculated as well. The simulation results are compared with analytical predictions showing an excellent agreement.
\end{abstract}

\section{Introduction}

An usual way of capturing the dissipative nature of granular media is through an idealized fluid of smooth, inelastic hard spheres. Despite the simplicity of the model, it has been shown to be quite useful in describing the dynamics of granular materials under rapid flow conditions \cite{C90,PL01}. The essential difference from ordinary fluids is the absence of energy conservation, leading to both obvious and subtle modifications of the Navier-Stokes hydrodynamic equations. In the context of kinetic theory, the Boltzmann and Enskog equations have been conveniently modified to account for inelastic binary collisions. These kinetic equations have been used to derive the corresponding fluid dynamic equations with explicit expressions for the transport coefficients for a monocomponent system \cite{BDKS98,GD99,GM02}, and for a binary mixture at low density \cite{GD02,MG03}. Recently, the shear viscosity has been also calculated for a dense binary mixture \cite{GM03}. In all these cases, the standard procedure to get the transport coefficients is the Chapman-Enskog expansion \cite{CC70} adapted to the case of inelastic collisions. 

In the case of elastic collisions, the Chapman-Enskog solution is obtained as an expansion around the local equilibrium distribution, while for inelastic collisions the reference state is a local {\em cooling} solution with a monotonically time decreasing temperature. For spatially homogeneous states, the latter is referred to as the {\em homogeneous cooling state} (HCS). Since the HCS qualifies as a normal solution, all its time dependence is only through the temperature. Nevertheless, in spite of its simplicity, no exact solution of the Boltzmann or Enskog equations describing such a state is known even for one-component systems \cite{NE98,MS00}. For a granular binary mixture, the set of coupled Enskog equations used to describe the system also admits a scaling solution in which the time dependence of the distribution functions occurs entirely through the temperature $T$ of the mixture. An important result is that the partial temperatures $T_i$ ($i=1,2$) of each species (measuring their mean kinetic energies) are {\em different}, although their cooling rates are equal  \cite{GD99bis,MG02}. This effect is generic for multicomponent systems and is a consequence of the inelasticity and the mechanical differences of the particles. This result contrasts with previous studies in granular mixtures \cite{JM89,Z95,AW98,WA99}, where the equality of the partial temperatures was assumed. The velocity distribution functions for each species are approximately determined by using a Sonine polynomial expansion around Maxwellians, which are defined in terms of the temperature for that species. The results show that, in general, the non-Maxwellian corrections have small effects on the cooling rates and on the temperature ratio even for strong dissipation. However, the corresponding reference Maxwellians for the two species are quite different due to temperature differences.  

As explained above, one of the main difficulties in obtaining the transport coefficients lies in the fact that, in contrast to what happens for elastic fluids, the reference state (HCS) depends on time due to the dissipation of energy through collisions. To overcome such difficulty, one possibility is to introduce external forces (thermostats) to accelerate the particles and hence compensate for collisional cooling. As a consequence, the corresponding reference state is stationary. This mechanism of energy input (different from those in shear flows or flows through vertical pipes) has been used by many authors in the past years to analyze different problems, such as non-Gaussian properties (cumulants, high-energy tails) of the velocity distribution function \cite{NE98,MS00}, long-range correlations \cite{NETP99}, and collisional statistics and short-scale structure \cite{PTNE02}. Since the latter requires the solution of the corresponding linearized hydrodynamic equations around the homogeneous state, the explicit expressions for the transport coefficients are needed. The use of Gauss's principle of least constraint leads to the Gaussian thermostat,  a force proportional to the particle velocity. For the sake of brevity, the resulting stationary state will be called here {\em heated stationary state} (HSS). It must be pointed out that the Enskog equations in the HSS  and in the HCS are formally identical when one scales the velocity with the thermal velocity \cite{MS00,GD99bis,MG02}. Therefore, the results obtained for the distribution functions and the temperature ratio $\gamma$ in the HCS applies to this thermostatted case as well. However, the response of the granular fluid when small spatial gradients are introduced is different in both cases. In particular, the shear viscosity calculated in the HCS does not coincide with that obtained in the HSS \cite{GM02,GM03}. The Navier-Stokes transport coefficients of a granular monocomponent gas have been obtained by solving the Enskog equation from the Chapman-Enskog method around the HSS \cite{GM02}. Also, the shear viscosity of a binary mixture in the HSS has been calculated for dilute \cite{MG03} and moderately dense \cite{GM03} systems. 

The partial temperatures and the velocity distribution functions are approximately determined in both the HCS and the HSS from a Sonine polynomial expansion \cite{MS00,MG02}. Once the HCS and the HSS are well characterized, the transport coefficients are given in terms of the solution of linear integral equations. These equations are solved by considering the leading terms in a Sonine polynomial expansion as well. A natural question is whether this type of approximation leads to accurate results. The direct simulation Monte Carlo (DSMC) method \cite{B94} has demonstrated to be an efficient and accurate procedure to analyse molecular gases far from equilibrium. This method, conveniently adapted to deal with inelastic collisions and finite density systems, provides an alternative route to study the behaviour of granular mixtures on the basis of the Boltzmann and Enskog equations, and so to establish the range of validity of the approximations used in the theoretical predictions mentioned above. The shear viscosity of a low-density granular gas in the HCS was determined from the DSMC method \cite{BRC99}. The results showed a very good agreement with the predictions based on the Boltzmann equation in the first Sonine approximation. The numerical experiment consisted in preparing an initial in-homogeneous nonequilibrium state corresponding to a transverse shear wave, and then analyzing its subsequent evolution in time. The shear wave decayed exponentially with a time scale inversely proportional to the viscosity. An alternative route for measuring the shear viscosity consists of preparing a state of uniform shear flow (USF) using Lees-Edwards boundary conditions \cite{LE72}. Macroscopically, this state is characterized by constant density, a uniform temperature, and a linear velocity profile. In a ordinary fluid, unless a thermostatting force is introduced, the temperature increases in time due to viscous heating. The corresponding energy balance equation can be used to determine the shear viscosity for sufficiently long times \cite{MS96,MS96b}. In a granular fluid, the relationship between the temperature and the shear viscosity is not so simple since there is a competition between viscous heating and collisional cooling. However, if external forces of the Gaussian form that exactly compensate for the collisional energy loss are introduced, the viscous heating effect is still able to heat the system. After a transient period, the system reaches a linear hydrodynamic regime where the shear viscosity can be calculated from the pressure tensor. This viscosity characterizes the momentum transport when a small perturbation is introduced in the HSS \cite{MG03,GM03}.   

The goal of this contribution is to present an over\-view of the results obtained by means of the Monte Carlo simulation for a binary granular mixture in both the HCS and the HSS. The numerical data for the temperature ratio, the velocity distribution functions, and the shear viscosity (in the case of the HSS) are compared with the analytical results showing an excellent agreement for the range of parameters considered. The plan of the paper is as follows. The theoretical framework and the procedure used to calculate the viscosity are described in Sec.\ \ref{sec2}. The details of the simulation method are given in Sec.\ \ref{sec3}. The analytical results and the simulation data are compared in Sec.\ \ref{sec4}. We close the paper in Sec.\ \ref{sec5} with a short summary and conclusions.

\section{Theoretical framework}
\label{sec2}

Consider a binary mixture of smooth hard spheres of masses $m_{1}$ and $m_{2}$ and diameters $\sigma_{1}$ and $\sigma_{2}$. The inelasticity of collisions among all pairs is characterized by three independent constant coefficients of normal restitution $\alpha_{11}$, $\alpha_{22}$, and $\alpha_{12}=\alpha_{21}$, where $\alpha_{ij}$ is the restitution coefficient for collisions between particles of species $i$ and $j$. The molar fractions $x_i\equiv n_i/n$ ($i=1,2$) indicate the relative concentration of the two species, while the density of the mixture can be measured by the volume packing fraction $\phi=\pi n/6\ (x_1\sigma_1^3+x_2\sigma_2^3)$. Here, $n_i$ is the number density corresponding to species $i$ and $n=n_1+n_2$. Of course, the monocomponent case is recovered when one considers mechanically equivalent particles, i.e., $m_1=m_2$, $\sigma_1=\sigma_2$ and $\alpha_{ij}=\alpha$. 

As mentioned in the Introduction, the viscosity characterizing the momentum transport in the HSS is calculated in our simulations by considering the USF with a thermostat force that exactly compensates for collisional cooling. The shear viscosity can be obtained from the ratio of the $xy$-element of the pressure tensor to the shear rate for sufficiently long times. This viscosity characterizes the evolution of the system when small spatial gradients of the velocity field (for instance, a transverse shear wave) are introduced in the HSS. Let us now introduce the Enskog equation under USF. At a microscopic level, the USF is generated by Lees-Edwards boundary conditions \cite{LE72}, which are simply periodic boundary conditions in the local Lagrangian frame ${\bf V}={\bf v}-{\sf a}\cdot {\bf r}$ and ${\bf R}={\bf r}-{\sf a}\cdot {\bf r}t$. Here, ${\sf a}$ is the tensor with elements $a_{\alpha\beta}=a\delta_{\alpha x}\delta_{\beta y}$, and $a$ is the shear rate. Using the above frame of reference, the USF becomes homogeneous. In terms of these variables, the velocity distribution functions $f_i({\bf V},t)$ ($i=1,2$) describing the granular fluid obey the equations
\begin{equation} 
\label{enskog}
\partial _{t}f_i-aV_y\frac{\partial}{\partial V_x} 
f_{i}+{\cal F}f_i=\sum_{j=1}^2J_{ij}^{\textin{E}}\left[ {\bf V}|f_{i}(t),f_{j}(t)\right] \;. 
\label{3.4} 
\end{equation} 
Here, ${\cal F}$ is an operator representing the effect of an external force (if it exists), and $J_{ij}^{\textin{E}}$ is the Enskog collision operator 
\begin{eqnarray} 
\label{3.5}
J_{ij}^{\textin{E}}\left[ {\bf V}_{1}|f_{i},f_{j}\right] &=&\sigma _{ij}^{2}\chi_{ij}\int d{\bf V} 
_{2}\int d\widehat{\bbox {\sigma }}\,\Theta (\widehat{{\bbox {\sigma }}} 
\cdot {\bf g})(\widehat{\bbox {\sigma }}\cdot {\bf g})  \left[ \alpha _{ij}^{-2} f_i({\bf V}_1',t)f_j({\bf V}_2',t)
-f_i( {\bf V}_1,t)f_j({\bf V}_2,t)\right]\; .
\end{eqnarray} 
In Eq.\ (\ref{3.5}), $\chi_{ij}$ is the pair correlation function for particles of type $i$ and $j$ when they are at contact, separated by $\sigma_{ij}=\left(\sigma_{i}+\sigma_{j}\right)/2$. Note that we have taken into account that $\chi_{ij}$ is uniform in all the states considered. In this case, the Carnahan-Starling approximation \cite{GH72} provides accurate results. It is given by
\begin{equation}
\label{4.6}
\chi_{ij}=\frac{1}{1-\phi}+\frac{3}{2}\frac{\widehat{\phi}\ \widetilde{\sigma}_{ij}}{(1-\phi)^{2}} +\frac{1}{2}\frac{(\widehat{\phi}\ \widetilde{\sigma}_{ij})^{2}}{(1-\phi )^{3}}\; ,
\end{equation}
where $\widehat{\phi}=(\pi n/6)\sum_{i}x_{i}\sigma_{ii}^{2}$ and $\widetilde{\sigma}_{ij}=\sigma_{ii}\sigma_{jj}/ \sigma_{ij}$. Also, in Eq.\ (\ref{3.5}) $\widehat{\bbox {\sigma}}$ is a unit vector along their line of centers, $\Theta$ is the Heaviside step function, ${\bf g}={\bf V}_1-{\bf V}_{2}-{\sf a}\cdot {\bbox \sigma}$, and
\begin{equation}
\label{sca1}
{\bf V}_{1}^{\prime }={\bf V}_{1}-\mu _{ji}\left(1+\alpha_{ij}^{-1}\right)
(\widehat{\bbox {\sigma}}\cdot {\bf g})\widehat{\bbox {\sigma}}\; ,
\end{equation}
\begin{eqnarray}
\label{sca2}
{\bf V}_{2}^{\prime}&=&{\bf V}_{2}+\mu_{ij}\left(1+\alpha_{ij}^{-1}\right) 
(\widehat{\bbox {\sigma}}\cdot {\bf g})\widehat{\bbox{\sigma}}\nonumber \\ 
&&+2a\ \sigma_{ij}\ \widehat{\sigma}_y\ \widehat{\bf x}\; ,
\end{eqnarray}
where $\mu_{ij}=m_{i}/\left( m_{i}+m_{j}\right)$. The collision operators (\ref{3.5}) conserve the particle number of each species and the total momentum, but the total energy is not conserved. The cooling rate $\xi$ is a measure of the loss of energy of the mixture due to collisions. In terms of the velocity distribution functions, it is given by 
\begin{eqnarray}
\label{3.7} 
\xi&=&\frac{1}{6nT}\sum_{i=1}^2\sum_{j=1}^2\frac{m_im_j}{m_i+m_j}\chi_{ij}\sigma_{ij}^2(1-\alpha_{ij}^2)
\int d{\bf V}_1\int d{\bf V}_2\int d\widehat{\bbox {\sigma }}\,\Theta (\widehat{{\bbox {\sigma}}} 
\cdot {\bf g})(\widehat{\bbox {\sigma }}\cdot {\bf g})^3  f_i\left({\bf V}_1,t\right)f_j({\bf V}_2,t)\; .
\end{eqnarray}

First, let us consider a system that evolves in the absence of shear. Obviously, the analysis can be also carried out from the above equations by simply setting $a=0$. In the case of elastic particles ($\alpha_{ij}=1$) and in the absence of external forcing (${\cal F}=0$), it is well-known that the long-time solution of Eq.\ (\ref{enskog}) is the Maxwell-Boltzmann equilibrium distribution function. On the other hand, if the particles are inelastic ($\alpha_{ij}<1$) and ${\cal F}=0$, a steady state is not possible in uniform situations since, due to the dissipation of energy through collisions, the temperature $T$ decreases monotonically with time. In this case, the solution of the Enskog equation tends to the HCS, characterized by the fact that all the time-dependence of $f_i$ occurs only through the temperature $T(t)$ of the mixture. Dimensional analysis requires that $f_i$ must be of the form
\begin{equation}
\label{2.11}
f_i(v;t)=n_iv_0^{-3}(t)\Phi_i(v^*)\; ,
\end{equation}
where $v_0(t)=\sqrt{2T(t)(m_1+m_2)/(m_1m_2)}$ is a thermal velocity defined in terms of the temperature of the mixture $T(t)$, and $v^*=v/v_0(t)$. Since the time dependence of $f_i$ occurs only through $T(t)$, it follows that the temperature ratio $\gamma\equiv T_1/T_2$ must reach a constant value in the long time limit. 

However, by driving a granular fluid by boundaries or external fields it can reach a steady state. The energy injected in the fluid may exactly compensate for the energy dissipated by collisional cooling. The same effect can be obtained by means of external forces (thermostats), ${\bf F}_i^{\textin{th}}$, acting locally on each particle. These forces are represented by the operator ${\cal F}$ in Eq.\ (\ref{enskog}). Several types of thermostat can be used. Here we consider the so-called {\em Gaussian} thermostat, i.e.,
\begin{equation}
\label{4.3}
{\bf F}_i^{\textin{th}}=\frac{1}{2}m_i\zeta{\bf V}, 
\end{equation}
where $\zeta$ is a positive constant. In this case, 
\begin{equation}
\label{ter}
{\cal F}f_i=\zeta \frac{\partial}{\partial {\bf V}}\cdot[{\bf V} f_i({\bf V})]\; .
\end{equation}
The steady state achieved by introducing this thermostat is the HSS. It must be pointed out that the Enskog equation in the HSS is formally identical with the Enskog equation in the HCS when one scales the velocity with the thermal velocity $v_0$ \cite{MS00,GD99bis,MG02}. Therefore, the results obtained for the distribution functions $\Phi_i$ and the temperature ratio $\gamma$ in the HCS applies to this thermostatted case as well. The same does not occur with the shear viscosity. 

In order to calculate the shear viscosity in the HSS, we assume now that the system is under USF. Let us consider first  the case of elastic collisions. In the absence of a thermostatting force, the energy balance equation yields the heating equation \cite{MS96,MS96b,NO79}
\begin{equation}
\label{4.1}
\partial_tT=-\frac{2}{3n}aP_{xy}\; , 
\end{equation}
where $P_{xy}$ is the $xy$ element of the stress tensor ${\sf P}$. Since the temperature $T$ increases in time, so does the collision frequency $\nu(t)\propto \sqrt{T(t)}$. As a consequence, the reduced shear rate $a^*(t)=a/\nu(t)$ (which is the relevant uniformity parameter) monotonically decreases in time and the system asymptotically tends towards that of equilibrium. This implies that for sufficiently long times (which means here $a^*\ll 1$) the system reaches a regime described by linear hydrodynamics, and the Navier-Stokes shear viscosity $\eta$ can be identified as \cite{NO79} 
\begin{equation}
\label{4.2}
\eta=-\lim_{t\to \infty}\frac{P_{xy}}{a}\; .
\end{equation}
This route has been shown to be quite efficient to measure the Navier-Stokes shear viscosity coefficient in ordinary fluids \cite{MS96,MS96b,Pepe}. For a granular mixture, unless a thermostat is introduced, the energy balance equation leads to a steady state when the viscous heating effect is exactly balanced by the collisional cooling \cite{C90,MGSB99,MG02a,MG03b,GM03b,CH02}. However, if the granular mixture is excited by the Gaussian thermostat that exactly compensates for the collisional energy loss (in this case $\zeta$ is the instantaneous value of the cooling rate $\xi$), the viscous heating still heats the system and Eq.\ (\ref{4.1}) remains valid. Consequently, the linear relationship (\ref{4.2}) allows one to determine the shear viscosity coefficient in the long time limit. It must be noted that the value of $\eta$ calculated in this way corresponds to the Navier-Stokes shear viscosity of an excited granular mixture (in the HSS) \cite{GM02,MG03,GM03}, and thus it does not necessarily coincide with the value obtained in the unforced case (in the HCS) \cite{GD02,BRC99}. 

In summary, the Enskog equations (\ref{enskog}) provide an adequate theoretical framework to study the granular properties in the HCS and in the HSS. If one takes $a=0$ and ${\cal F}=0$, then the system evolves to the HCS independently of the initial conditions considered. For $a=0$ and ${\cal F}$ given by (\ref{ter}), the HSS is achieved. In order to obtain the shear viscosity coefficient $\eta$ of the mixture in this state, an arbitrary value $a\neq 0$ has to be taken (USF) and the thermostat force (\ref{ter}) has to be introduced with $\zeta$ equal to the instantaneous value of the cooling rate $\xi$. The coefficient $\eta$ can be measured in the long time limit.

As mentioned in the Introduction, in this context the theoretical predictions are determined by means of a Sonine polynomial expansion. In the next section a Monte Carlo algorithm devised to solve (\ref{enskog}) is described. The numerical results obtained are compared with the analytical ones in Sec.\ \ref{sec4}.

\section{Monte Carlo simulation method}
\label{sec3}

The Enskog Simulation Monte Carlo (ESMC) method \cite{MS96,MS96b} is an extension of the well-known direct simulation Monte Carlo (DSMC) method \cite{B94} to dense gases, and it was devised to mimic the dynamics involved in the Enskog collision term. Here we briefly describe the adaptation of this procedure to numerically solve the problem (\ref{enskog}). It must be pointed out that since the sought solutions are spatially homogeneous (in the case of the USF thanks to the Lagrangian frame used), the simulation method becomes especially easy to implement and efficient from a computational point of view. This is an important advantage with respect to molecular dynamics simulations. Nevertheless, the restriction to this homogeneous states prevents us from analyzing the possible instability of the USF or the formation of clusters or microstructures. 

The ESMC method as applied to (\ref{enskog}) is as follows. The velocity distribution function of the species $i$ is represented by the peculiar velocities $\{{\bf V}_k\}$ of $N_i$ ``simulated" particles:
\begin{equation}
\label{4.190}
f_i({\bf V},t)\to n_i \frac{1}{N_i}\sum_{k=1}^{N_i} \delta({\bf V}-{\bf V}_k(t))\; .
\end{equation}
Note that the number of particles $N_i$ must be taken according to the relation $N_1/N_2=n_1/n_2$. At the initial state, one assigns velocities to the particles drawn from the Maxwell-Boltzmann probability distribution: 
\begin{equation}
\label{4.2b}
f_i({\bf V},0)=n_i\ \pi^{-3}\ V_{0i}^{-3}(0)\ \exp\left(-V^2/V_{0i}^2(0)\right)\;,
\end{equation}
where $V_{0i}^2(0)=2T(0)/m_i$ and $T(0)$ is the initial temperature. To enforce a vanishing initial total momentum, the velocity of every particle is subsequently subtracted by the amount $N_i^{-1} \sum_k {\bf V}_k(0)$. In the ESMC method, the free motion and the collisions are uncoupled over a time step $\Delta t$ which is small compared with the mean free time and the inverse reduced shear rate $a^{*-1}$ (if $a\neq 0$). If those quantities vary significantly, the value of $\Delta t$ must be conveniently updated in the course of the simulation. 

If $a\neq 0$, in the local Lagrangian frame particles of each species ($i=1,2$) are subjected to the action of a non-conservative inertial force ${\bf F}_i=-m_i\ {\sf a}\cdot{\bf V}$. Thus, the free motion stage consists of making ${\bf V}_k\to {\bf V}_k-{\sf a}\cdot{\bf V}_k\Delta t$. In the collision stage, binary interactions between particles of species $i$ and $j$ must be considered. To simulate the collisions between particles of species $i$ with $j$ a sample of $\frac{1}{2} N_i \omega_{\textin{max}}^{(ij)}\Delta t$ pairs is chosen at random with equiprobability. Here, $\omega_{\textin{max}}^{(ij)}$ is an upper bound estimate of the probability that a particle of the species $i$ collides with a particle of the species $j$. Let us consider a pair $\{k,\ell\}$ belonging to this sample. Hereafter, $k$ denotes a particle of species $i$ and $\ell$ a particle of species $j$. For each pair $\{k,\ell\}$ with velocities $\{{\bf V}_k,{\bf V}_{\ell}\}$, the following steps are taken: (1) a given direction $\widehat{\bbox \sigma}_{k\ell}$ is chosen at random with equiprobability; (2) the collision between particles $k$ and $\ell$ is accepted with a probability equal to $\Theta({\bf g}_{k\ell}\cdot \widehat{\bbox \sigma}_{k\ell})\omega_{k\ell}^{(ij)}/ \omega_{\textin{max}}^{(ij)}$, where $\omega_{k\ell}^{(ij)}=4\pi \sigma_{ij}^2 n_j|{\bf g}_{k\ell}\cdot \widehat{\bbox \sigma}_{k\ell}|$ and ${\bf g}_{k\ell}={\bf V}_k-{\bf V}_{\ell}-\sigma_{ij} {\sf a}\cdot \widehat{\bbox \sigma}_{k\ell}$; (3) if the collision is accepted, postcollisional velocities are assigned to both particles according to the counterparts of the rule (\ref{sca1}):
\begin{equation}
\label{4.3b}
{\bf V}_{k}\to {\bf V}_{k}-\mu_{ji}(1+\alpha_{ij})({\bf g}_{k\ell}\cdot \widehat{\bbox 
\sigma}_{k\ell})\widehat{\bbox \sigma}_{k\ell}\; ,
\end{equation}
\begin{equation}
\label{4.4b}
{\bf V}_{\ell}\to {\bf V}_{\ell}+\mu_{ij}(1+\alpha_{ij})({\bf g}_{k\ell}\cdot \widehat{\bbox 
\sigma}_{k\ell})\widehat{\bbox \sigma}_{k\ell}\;.
\end{equation}
If in a collision $\omega_{k\ell}^{(ij)}>\omega_{\textin{max}}^{(ij)}$, the estimate of $\omega_{\textin{max}}^{(ij)}$ is updated as $\omega_{\textin{max}}^{(ij)}=\omega_{k\ell}^{(ij)}$. The procedure described above is performed for $i=1,2$ and $j=1,2$. 

In the HSS case, after the collisions have been calculated, the thermostat force (\ref{4.3}) is considered by making 
\begin{equation}
\label{ters}
{\bf V}_k\to {\bf V}_k+1/2\ \zeta {\bf V}_k\Delta t\; .
\end{equation}
To obtain the shear viscosity value in this state, an arbitrary value $a\neq 0$ has to be taken (USF), and the constant $\zeta$ in (\ref{ters}) must be equal to the instantaneous value of the cooling rate $\xi$. The latter is obtained by calculating the granular temperature before and after the collision stage. To get the rest of quantities in the HSS, $a$ is set equal to zero and $\zeta$ can be chosen arbitrarily. In fact, this choice only affects to the value of the granular temperature in the steady state.

In the course of the simulations, one evaluates the distribution function of the species $i$, its moments and the partial temperatures $T_i$ in the usual way. In addition, the kinetic and collisional transfer contributions to the pressure tensor are calculated by the expressions
\begin{equation}
\label{4.5}
{\sf P}^{\textin{k}}=\sum_{i=1}^{2} \frac{m_i n_i}{N_i}\sum_{k=1}^{N_i} {\bf V}_k {\bf V}_k\; ,
\end{equation}
\begin{equation}
{\sf P}^{\textin{c}}=\frac{n}{2N\Delta t}\sum_{k\ell}^{\dagger} \mu_{ij}m_j \sigma_{ij}(1+\alpha_{ij})
({\bf g}_{k\ell}\cdot \widehat{\bbox \sigma}_{k\ell})\widehat{\bbox \sigma}_{k\ell}
\widehat{\bbox \sigma}_{k\ell}\; ,
\end{equation}
where the dagger means that the summation is restricted to the accepted collisions. The shear viscosity $\eta$ is obtained from (\ref{4.2}), while its kinetic contribution is $\eta^{\textin{k}}=-\lim_{t\to \infty}P_{xy}^{\textin{k}}/a$. In our simulations we have typically taken a total number of particles $N=N_1+N_2=10^5$ and a time step $\Delta t=3\times 10^{-3} \lambda_{11}/V_{01}(0)$, where $\lambda_{11}=(\sqrt{2} \pi n_1 \sigma_{11}^2)^{-1}$ is the mean free path for collisions 1--1. To improve the statistics, the results are averaged over a number ${\cal N}=10$ of independent realizations or replicas.

\section{Results}
\label{sec4}

In this section we compare the analytical predictions mentioned in the Introduction with the numerical results obtained from the simulation procedure described above. The dimensionless quantities calculated in both theory and simulation are functions of the set of parameters $\{\mu\equiv m_1/m_2$, $\omega\equiv \sigma_1/\sigma_2$, $\delta\equiv n_1/n_2$, $\alpha_{ij}$, $\phi\}$. For the sake of simplicity, in what follows we shall assume that $\alpha_{11}=\alpha_{22}= \alpha_{12} \equiv \alpha$.

\begin{figure}[hbt]
\begin{center}\epsfig{file=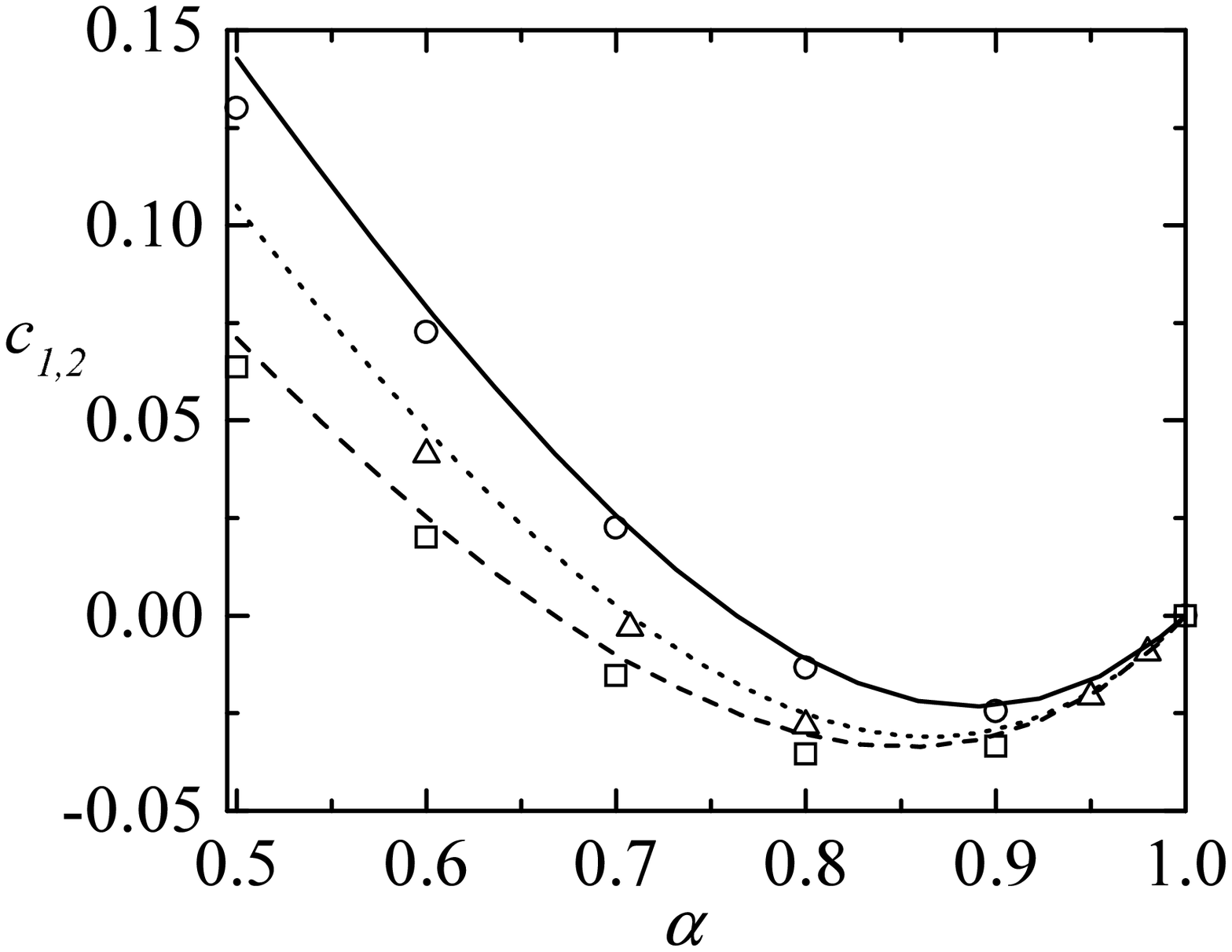,width=6.5cm}\end{center}
\caption{Plot of the coefficients $c_i$ versus the restitution coefficient $\alpha$ for $\mu=2$, $\omega=\delta=1$ and $\phi=0$. The solid line and the circles refer to $c_1$ while the dashed line and the squares correspond to $c_2$. The dotted line and the triangles refer to the common value in the single component case. The lines are the theoretical predictions and the symbols correspond to the simulation results.}
\label{fig1}
\end{figure}

\begin{figure}[hbt]
\begin{center}\epsfig{file=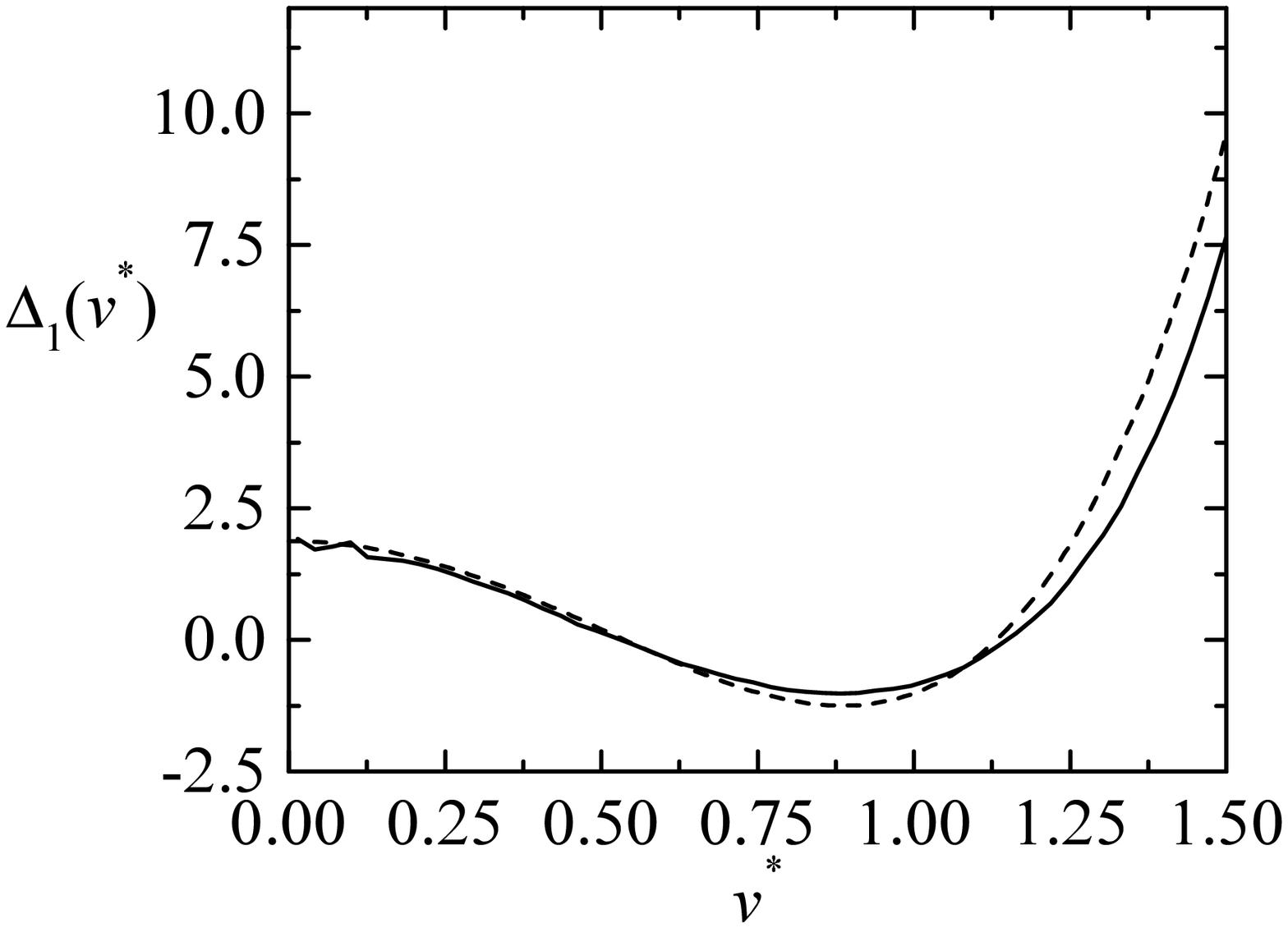,width=6.5cm}\end{center}
\caption{Plot of the simulation values of the function $\Delta_1(v^*)$ defined by Eq.\ (\protect\ref{5.1}) for 
$\mu=4$, $\omega=1$, $\delta=1/2$, $\alpha=0.5$ and $\phi=0$. The dashed line is the Sonine polynomial (\protect\ref{5.2}).}
\label{fig2}
\end{figure}

Let us consider first the analysis in the absence of shear rate. The basic quantity measuring the deviation of the distribution functions from the corresponding Maxwellians are the cumulants $c_i$ \cite{MS00,MG02}. In Fig.\ \ref{fig1}, we show the dependence of $c_1$ and $c_2$ on $\alpha$ for $\mu=2$, $\omega=1$, $\delta=1$ and $\phi=0$. We also present the corresponding result for the one-component system (mechanically equivalent particles, i.e., $\mu=\omega=1$). Note that the results apply in both the HCS and in the HSS. The agreement between the simulation data and the theoretical predictions is excellent. The small values of these coefficients support the assumption of a low-order truncation in polynomial expansion and indicate that the distribution functions for thermal velocities are well represented by the Sonine approximation. To confirm this, we have measured the deviation of the distribution functions $\Phi_i$ given by (\ref{2.11}) from the corresponding Maxwellian. More precisely, we have evaluated the function $\Delta_i(v^*)$ defined by the relation 
\begin{equation}
\label{5.1}
\Phi_i(v^*)=\left(\frac{\lambda_i}{\pi}\right)^{3/2}e^{-\lambda_i v^{*2}} \left[1+\case{1}{2}c_i\Delta_i(v^*)\right]\; ,
\end{equation}
where $\lambda_i=T/(T_i\mu_{ji})$. The function $\Delta_1(v^*)$ is plotted in Fig.\ \ref{fig2} for $\mu=4$, $\omega=1$, $\delta=1/2$, $\alpha=0.5$ and $\phi=0$. The dashed line is the first Sonine approximation
\begin{equation}
\label{5.2} 
\Delta_1(v^*)\to \case{1}{2} \lambda_i^2 v^{*4}-\case{5}{2}\lambda_iv^{*2}+\case{15}{8}\; .
\end{equation} 
As could be expected, the simulation curve agrees very well with the corresponding Sonine polynomial, confirming the accuracy of the analytical solution in the region of thermal velocities.  

\begin{figure}[hbt]
\begin{center}\epsfig{file=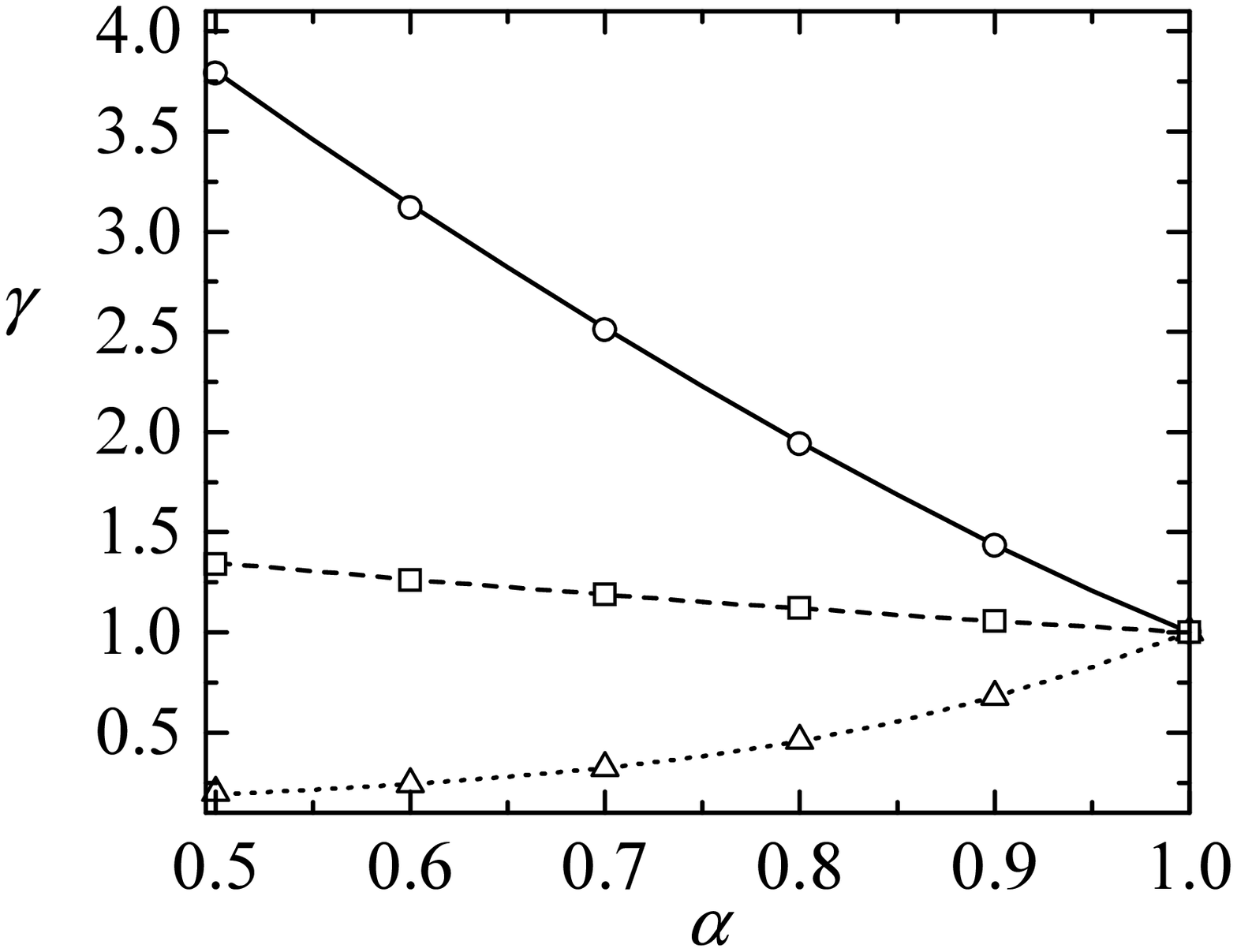,width=6.5cm}\end{center}
\caption{Plot of the temperature ratio $\gamma$ versus the restitution coefficient $\alpha$ for $\omega=1$, $\delta=2$, $\phi=0$ and three different values of the mass ratio: $\mu=1/10$ (dotted line and triangles), $\mu=2$ (dashed line and squares) and $\mu=10$ (solid line and circles). The lines are the theoretical predictions and the symbols correspond to the simulation results.}
\label{fig3}
\end{figure}

\begin{figure}[hbt]
\begin{center}\epsfig{file=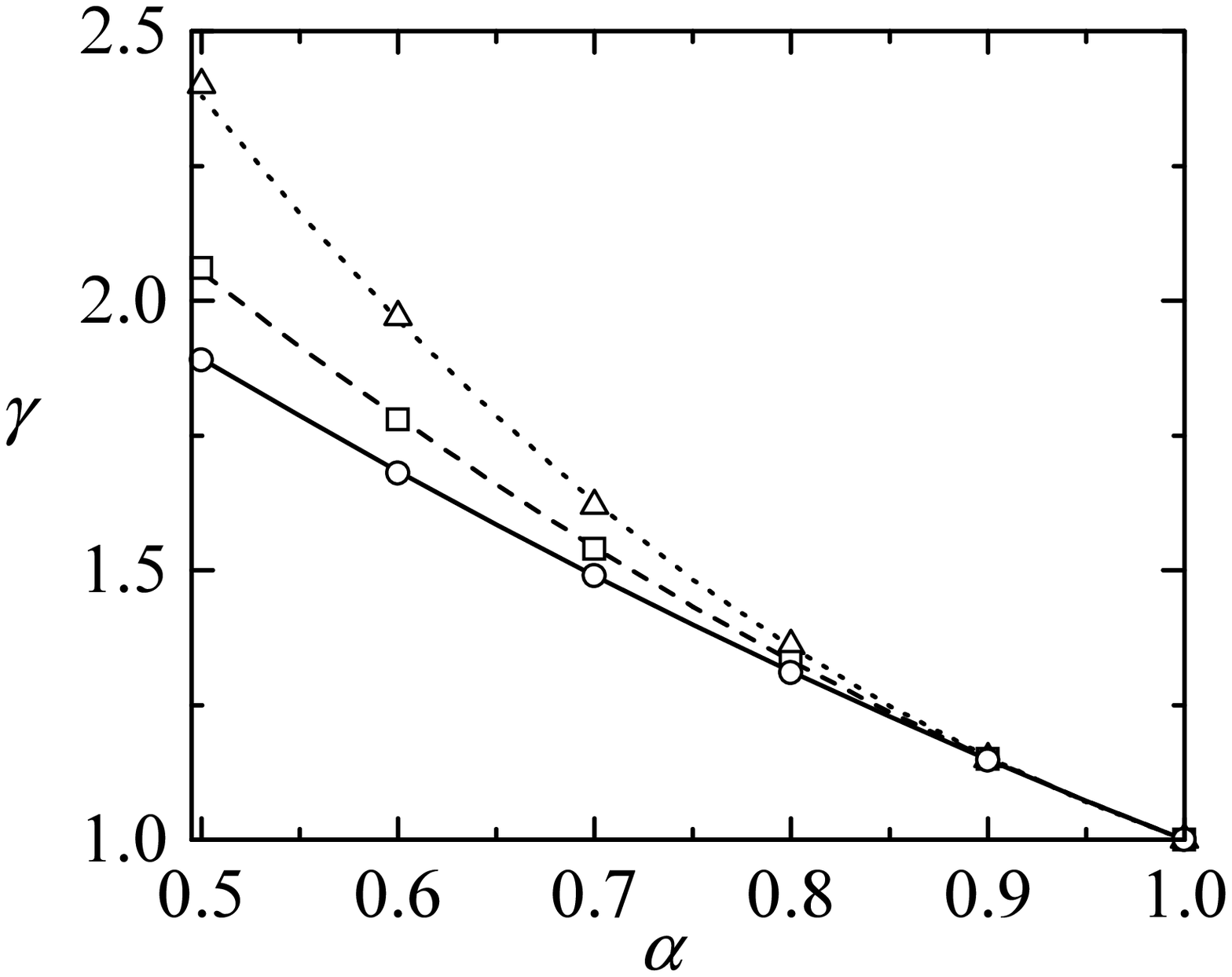,width=6.5cm}\end{center} 
\caption{Plot of the temperature ratio versus the restitution coefficient $\alpha$ for $\mu=4$, $\omega=1$, $\phi=0$ and three different values of the concentration ratio: $\delta=1/4$ (dotted line and triangles), $\delta=1$ (dashed line and squares) and $\delta=4$ (solid line and circles). The lines are the theoretical predictions and the symbols correspond to the simulation results.}
\label{fig4}
\end{figure}

\begin{figure}[hbt]
\begin{center}\epsfig{file=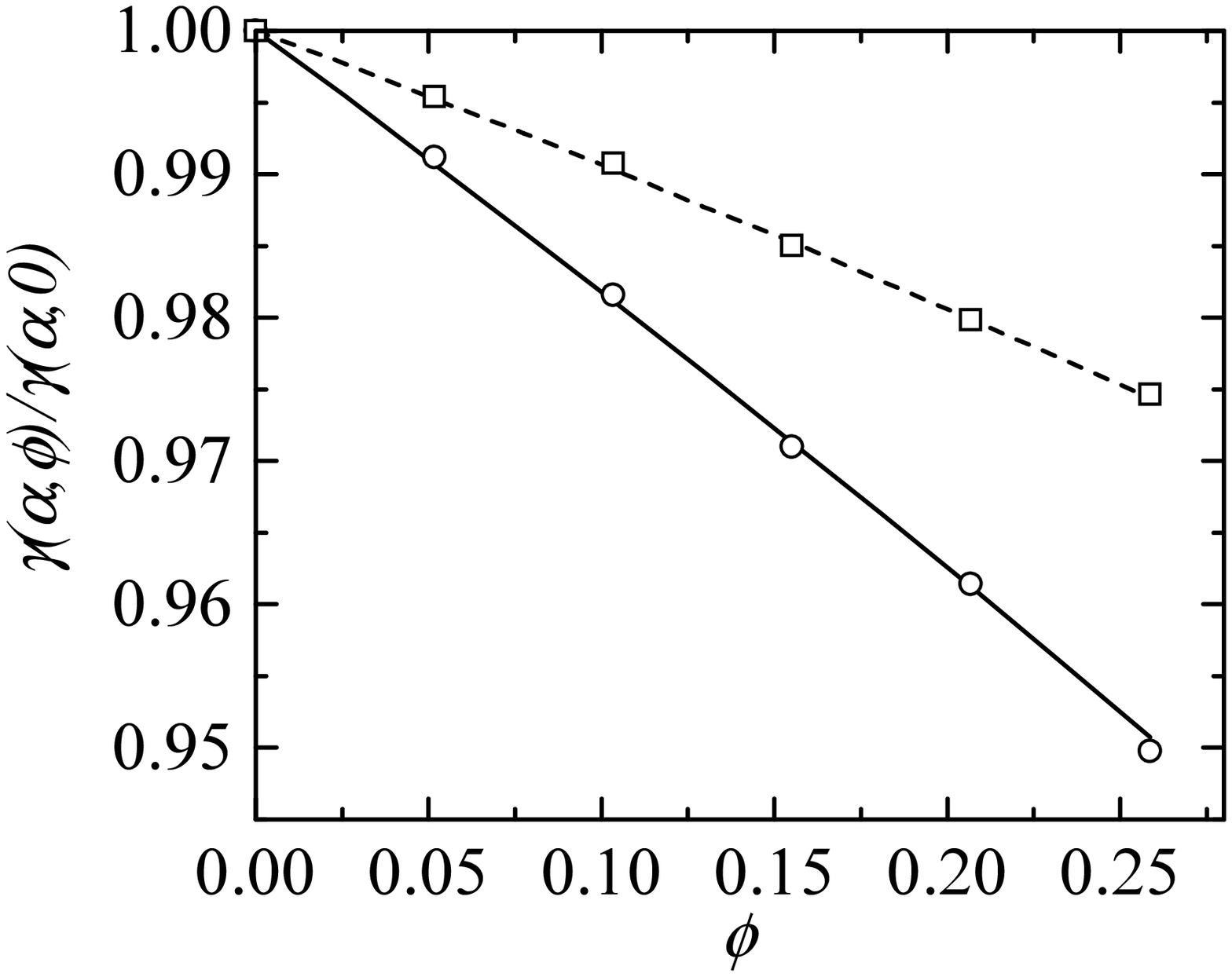,width=6.5cm}\end{center} 
\caption{Plot of the relative temperature ratio $\gamma(\alpha,\phi)/\gamma(\alpha,0)$ as a function of the volume packing fraction $\phi$ for $\mu=\omega=2$, $\delta=1/2$, and two different values of the restitution coefficient: $\alpha=0.6$ (solid line and circles) and $\alpha=0.8$ (dashed line and squares). The lines are the theoretical predictions and the symbols correspond to the simulation results.}
\label{fig5}
\end{figure}

One of the main new results of the description showed in Refs.\ \cite{GD99bis,MG02} is that the partial temperatures are different ($\gamma\neq 1$). This conclusion contrasts with previous results derived for granular mixtures \cite{JM89,Z95,AW98,WA99}, where it was implicitly assumed the equipartition of granular energy between both species (i.e., $\gamma=1$). In Figs.\ \ref{fig3}, \ref{fig4}, and \ref{fig5} we plot the temperature ratio $\gamma$ versus the restitution coefficient $\alpha$ for different choices of the mechanical parameters characterizing the mixture. The theory as well as the simulation results clearly indicate that $\gamma$ is different from unity, even for weak inelasticity. Figure \ref{fig3} shows the dependence of $\gamma$ on $\alpha$ for $\omega=1$, $\delta=2$, $\phi=0$ and several values of the mass ratio. The agreement between theory and simulation is very good, implying the accuracy of the expression of $\gamma$ obtained in the first Sonine approximation. For large differences in the mass ratio, the temperature differences are significant. As can be observed in Fig.\ \ref{fig4}, the influence of the concentration ratio $\delta$ on the temperature ratio $\gamma$ is not as strong as that observed with the mass ratio, although is still quite important. The dependence of the relative temperature ratio $\gamma(\alpha,\phi)/\gamma(\alpha,0)$ on the volume packing fraction $\phi$ is plotted in Fig.\ \ref{fig5} for $\mu=2$, $\omega=2$, $\delta=1/2$, and two different values of $\alpha$: $\alpha=0.6$ and $0.8$. We see that, for a given value of the density, the relative temperature ratio decreases as the degree of inelasticity increases. It is evident again the excellent agreement between the Sonine predictions and the simulation data.

\begin{figure}[hbt]
\begin{center}\epsfig{file=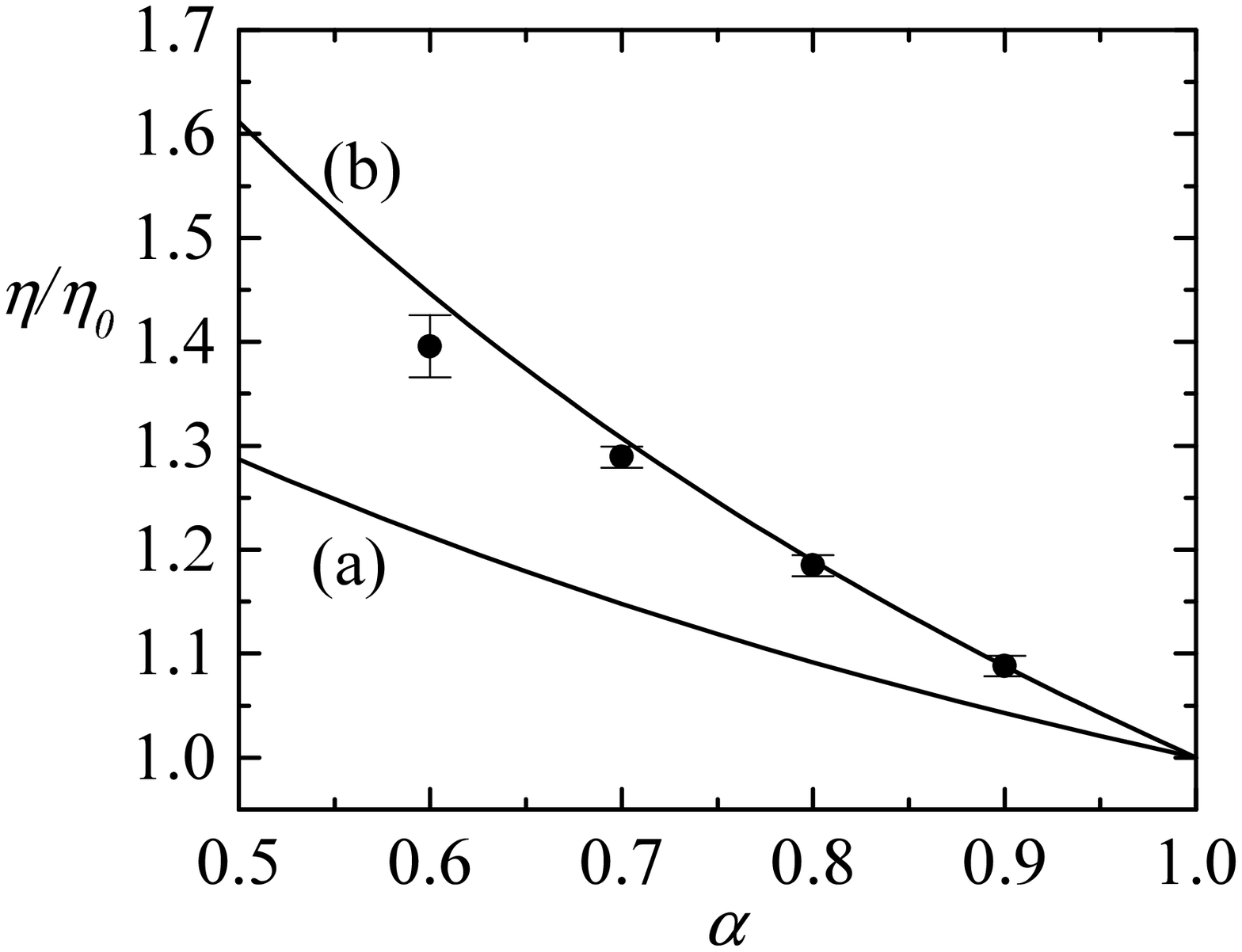,width=6.5cm}\end{center}
\caption{Reduced shear viscosity $\eta/\eta_0$ as a function of the restitution coefficient $\alpha$ for a monocomponent gas at low density ($\phi=0$). The solid lines refer to the analytical results derived in the HCS (a) and in the HSS (b). The symbols are the results obtained from the simulation in the HSS.}
\label{fig6}
\end{figure}

Let us focus on the sudy of the shear viscosity coefficient $\eta$. The theoretical analysis carried out in Refs.\ \cite{GM02,MG03,GM03} shows that the transport properties are affected by the thermostat introduced. In particular, the values obtained for the shear viscosity in the HCS and in the HSS are different. In Fig.\ \ref{fig6} we plot the reduced shear viscosity $\eta/\eta_0$, $\eta_0$ being the corresponding elastic value, in the HCS and in the HSS for a monocomponent gas. In both cases the viscosity increases with the dissipation. However, at a quantitative level, the influence of dissipation on the viscosity in the HCS case is smaller than in the HSS. The agreement between theory and simulation is quite good. The same trends are observed for a binary mixture (cf.\ Fig.\ 1 in Ref.\ \cite{GM03}). 

\begin{figure}[hbt]
\begin{center}\epsfig{file=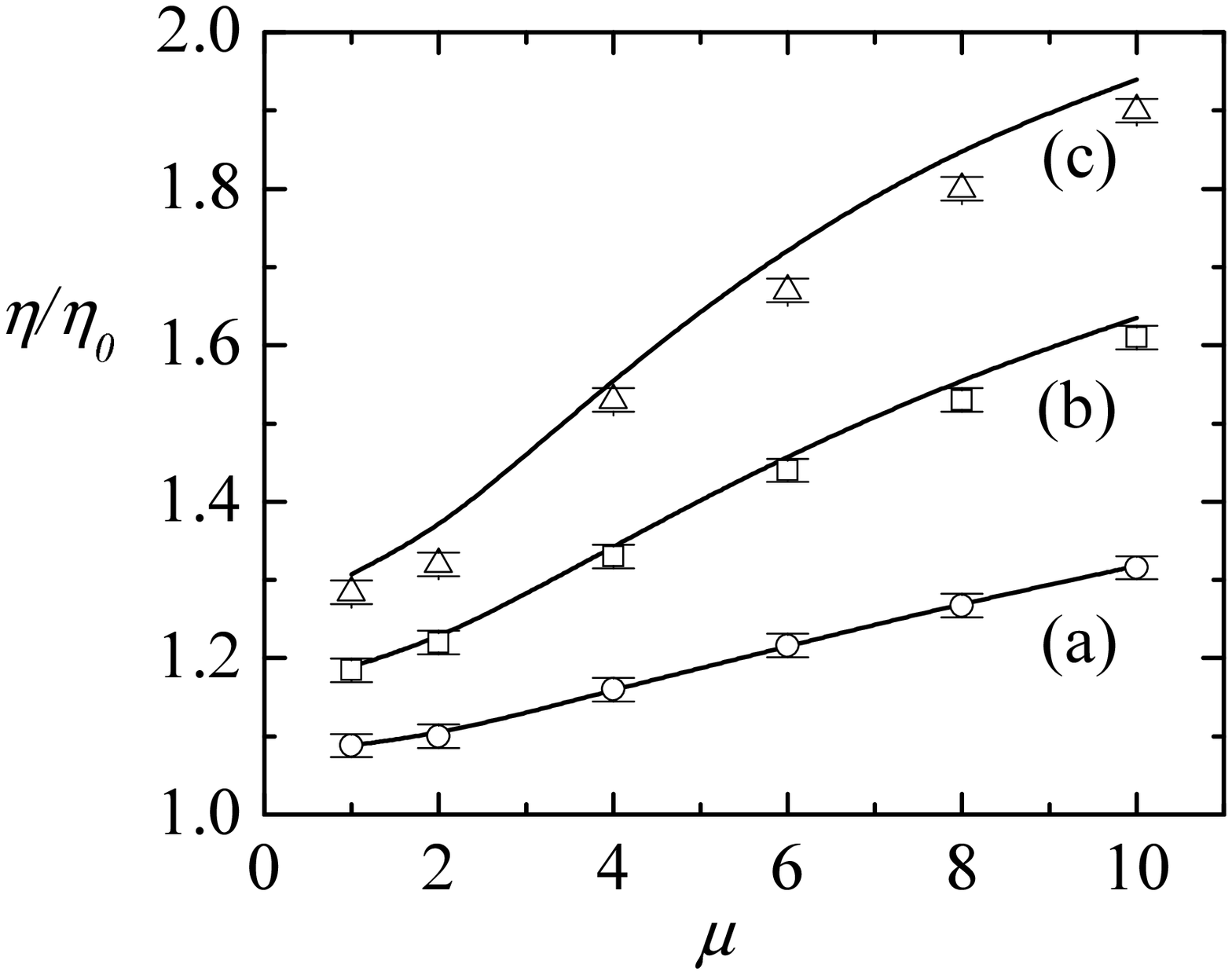,width=6.5cm}\end{center} 
\caption{Plot of the ratio $\eta/\eta_0$ as a function of the mass ratio $\mu$ for $\omega=\delta=1$ and three different values of the restitution coefficient $\alpha$: (a) $\alpha=0.9$ (circles), (b) $\alpha=0.8$ (squares), and (c) $\alpha=0.7$ (triangles). The lines are the theoretical predictions and the symbols refer to the simulations.}
\label{fig7}
\end{figure}

\begin{figure}[hbt]
\begin{center}\epsfig{file=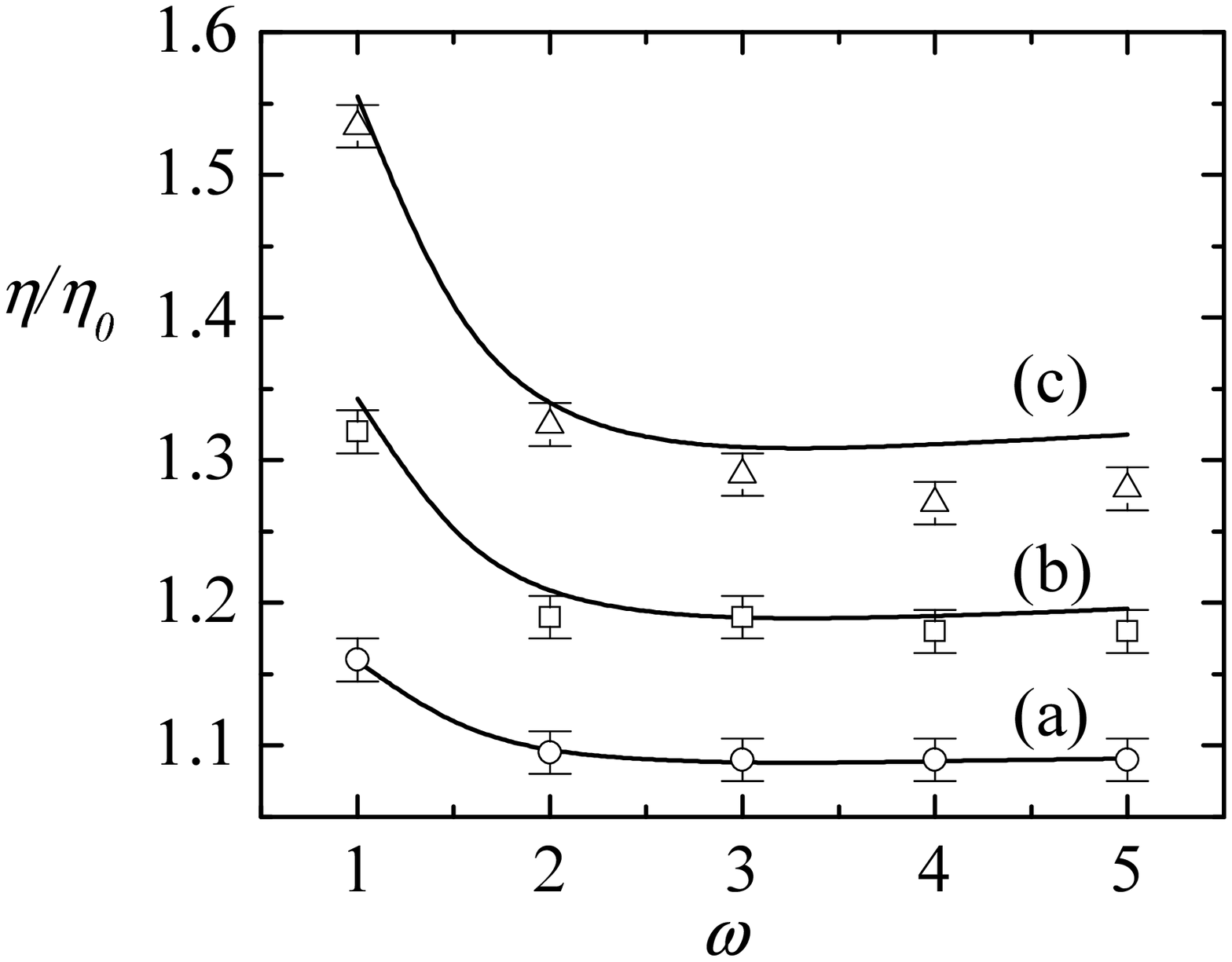,width=6.5cm}\end{center} 
\caption{Plot of the ratio $\eta/\eta_0$ as a function of the size ratio $\omega$ for $\mu=4$, $\delta=1$ and three different values of the restitution coefficient $\alpha$: (a) $\alpha=0.9$ (circles), (b) $\alpha=0.8$ (squares), and  (c) $\alpha=0.7$ (triangles). The lines are the theoretical predictions and the symbols refer to the simulations.}
\label{fig8}
\end{figure}

\begin{figure}[hbt]
\begin{center}\epsfig{file=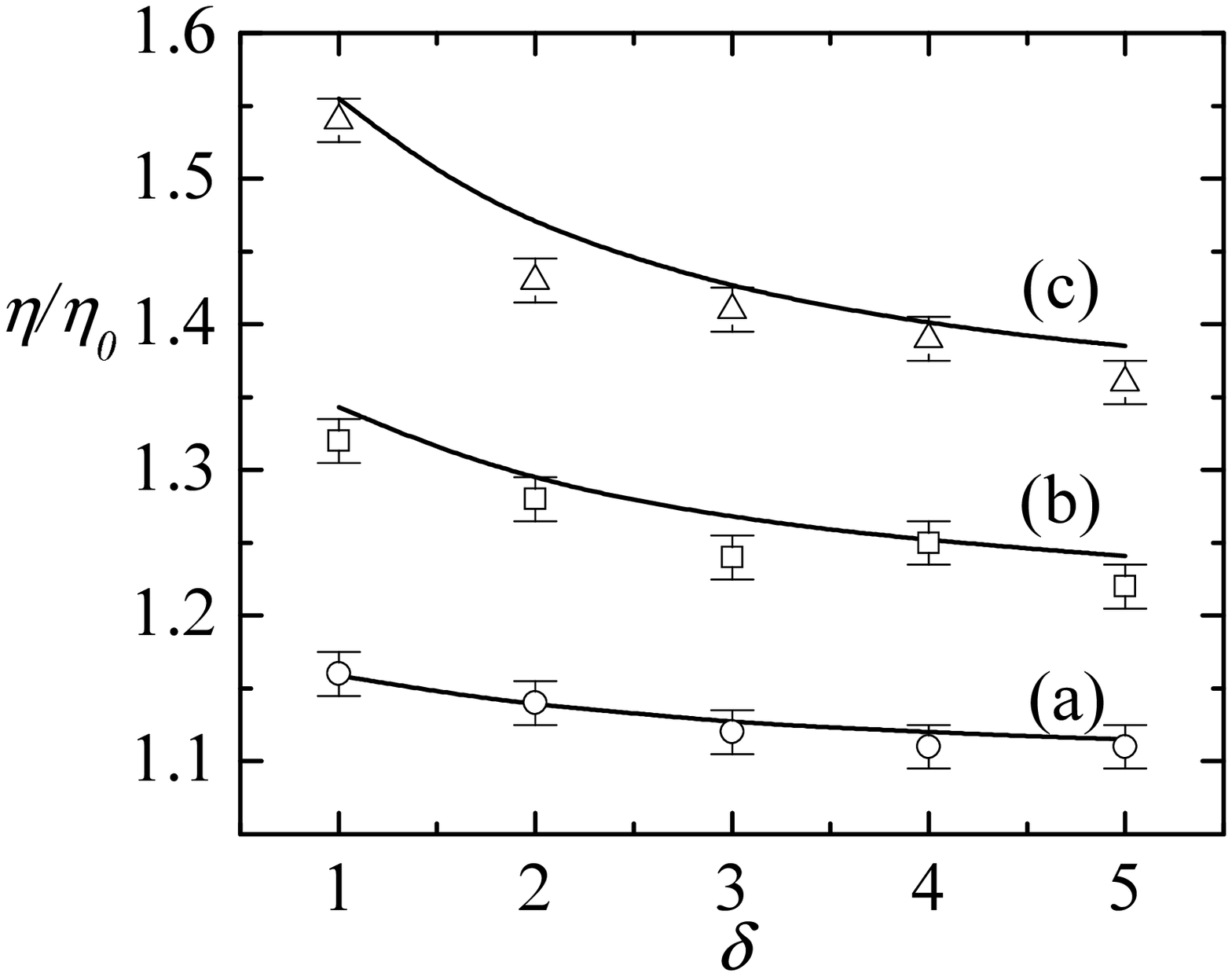,width=6.5cm}\end{center}
\caption{Plot of the ratio $\eta/\eta_0$ as a function of the concentration ratio $\delta$ for $\mu=4$, $\omega=1$ and three different values of the restitution coefficient $\alpha$: (a) $\alpha=0.9$ (circles), (b) $\alpha=0.8$ (squares), and (c) $\alpha=0.7$ (triangles). The lines are the theoretical predictions and the symbols refer to the simulations.}
\label{fig9}
\end{figure}

Next, in Figs.\ \ref{fig7}, \ref{fig8} and \ref{fig9} we show the influence of dissipation on the reduced shear viscosity $\eta(\alpha)/\eta_0$ for a heated granular binary mixture at low density ($\phi=0$) with different values of the mass ratio, the size ratio, and the concentration ratio. Three different values of the restitution coefficient are considered: $\alpha=0.9, 0.8$, and $0.7$. In Fig.\ \ref{fig7}, we plot the ratio $\eta(\alpha)/\eta_0$ versus the mass ratio $\mu$ for $\omega=\delta=1$. Again, the symbols represent the simulation data while the lines refer to the theoretical results obtained from the Boltzmann equation in the first Sonine approximation. We see that in general the deviation of $\eta(\alpha)$ from its functional form for elastic collisions is quite important. This tendency becomes more significant as the mass disparity increases. The agreement between the first Sonine approximation and simulation is seen to be in general quite good. This agreement is similar to the one found in the monocomponent case (Fig.\ \ref{fig6}). At a quantitative level, the discrepancies between theory and simulation tend to increase as the restitution coefficient decreases, although these differences are quite small (say, for instance, around 2\% at $\alpha=0.7$ in the disparate mass case $m_1/m_2=10$). 

The influence of the size ratio on the shear viscosity is shown in Fig.\ \ref{fig8} for $\mu=4$ and $\delta=1$. We observe again a strong dependence of the shear viscosity on dissipation. However, for a given value of $\alpha$, the influence of $\omega$ on $\eta$ is weaker than the one found before in Fig.\ \ref{fig7} for the mass ratio. The agreement for both $\alpha=0.9$ and $\alpha=0.8$ is quite good, except for the largest size ratio at $\alpha=0.8$. These discrepancies become more significant as the dissipation increases (say, for instance, $\alpha=0.7$), especially for mixtures of particles of very different sizes. Finally, Fig.\ \ref{fig9} shows the dependence of $\eta(\alpha)/\eta_0$ on the concentration ratio for $\mu=4$ and $\omega=1$. We observe that both the theory and simulation predict a very weak influence of composition on the shear viscosity. With respect to the influence of dissipation, the trends are similar to those found in Figs.\ \ref{fig7} and \ref{fig8}: the main effect of inelasticity in collisions is to enhance the momentum transport with respect to the case of elastic collisions. The agreement now between theory and simulation is very good, even for disparate values of the concentration ratio and/or strong dissipation. Therefore, according to the comparison carried out in Figs.\ \ref{fig7}, \ref{fig8}, and \ref{fig9}, we can conclude that the agreement extends over a wide range of values of the restitution coefficient, indicating the reliability of the first Sonine approximation for describing granular flows beyond the quasielastic limit. 

\begin{figure}[hbt]
\begin{center}\epsfig{file=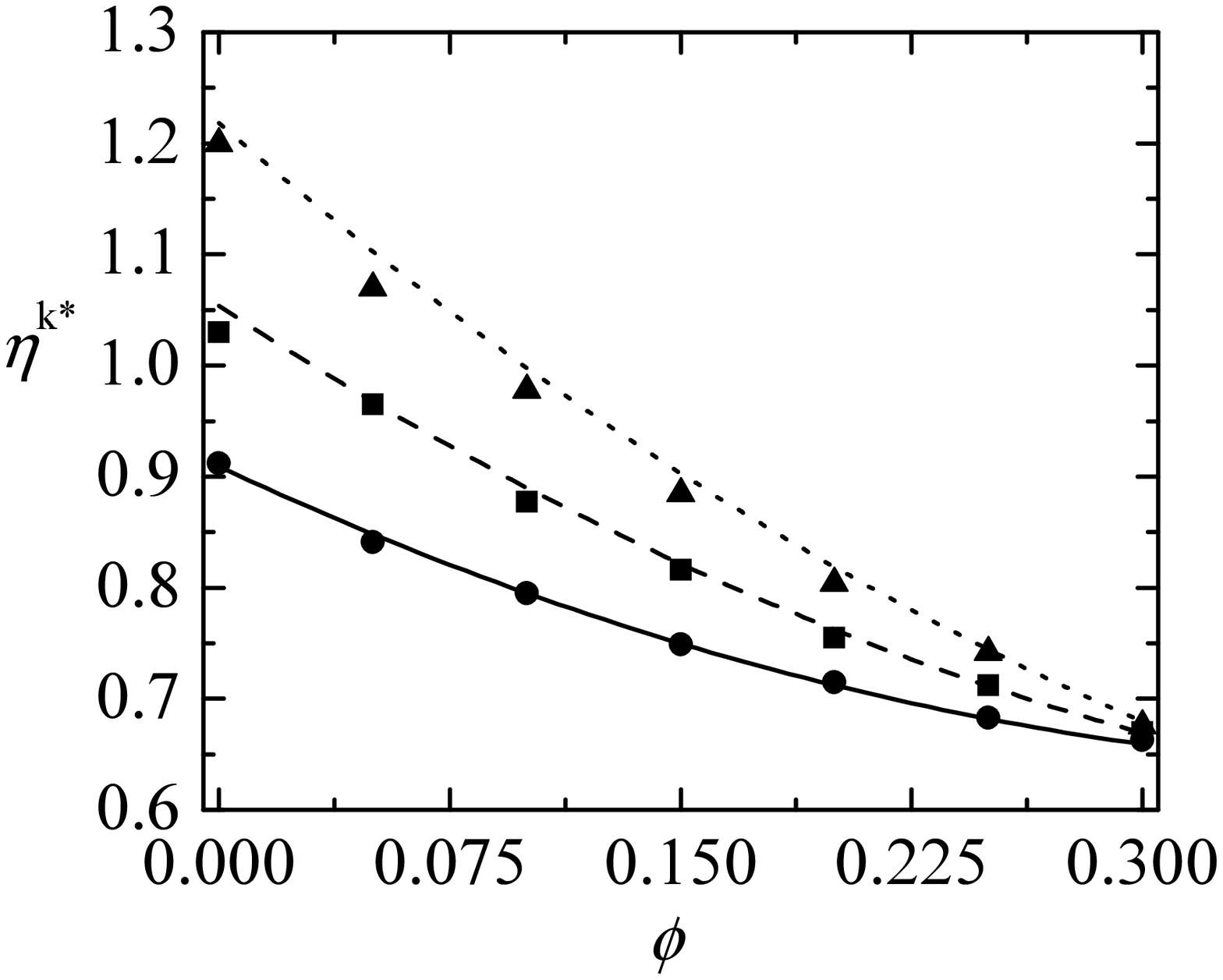,width=6.5cm}\end{center}
\caption{Plot of the kinetic part $\eta^{\textin{k}*}$ of the reduced shear viscosity as a function of the solid fraction $\phi$ for $\mu=4$, $\omega=\delta=1$ and three different values of the restitution coefficient $\alpha$: $\alpha=0.9$ (solid line and circles), $\alpha=0.8$ (dashed line and squares), and $\alpha=0.7$ (dotted line and triangles). The lines are the theoretical predictions and the symbols refer to the simulations.}
\label{fig10}
\end{figure}

\begin{figure}[hbt]
\begin{center}\epsfig{file=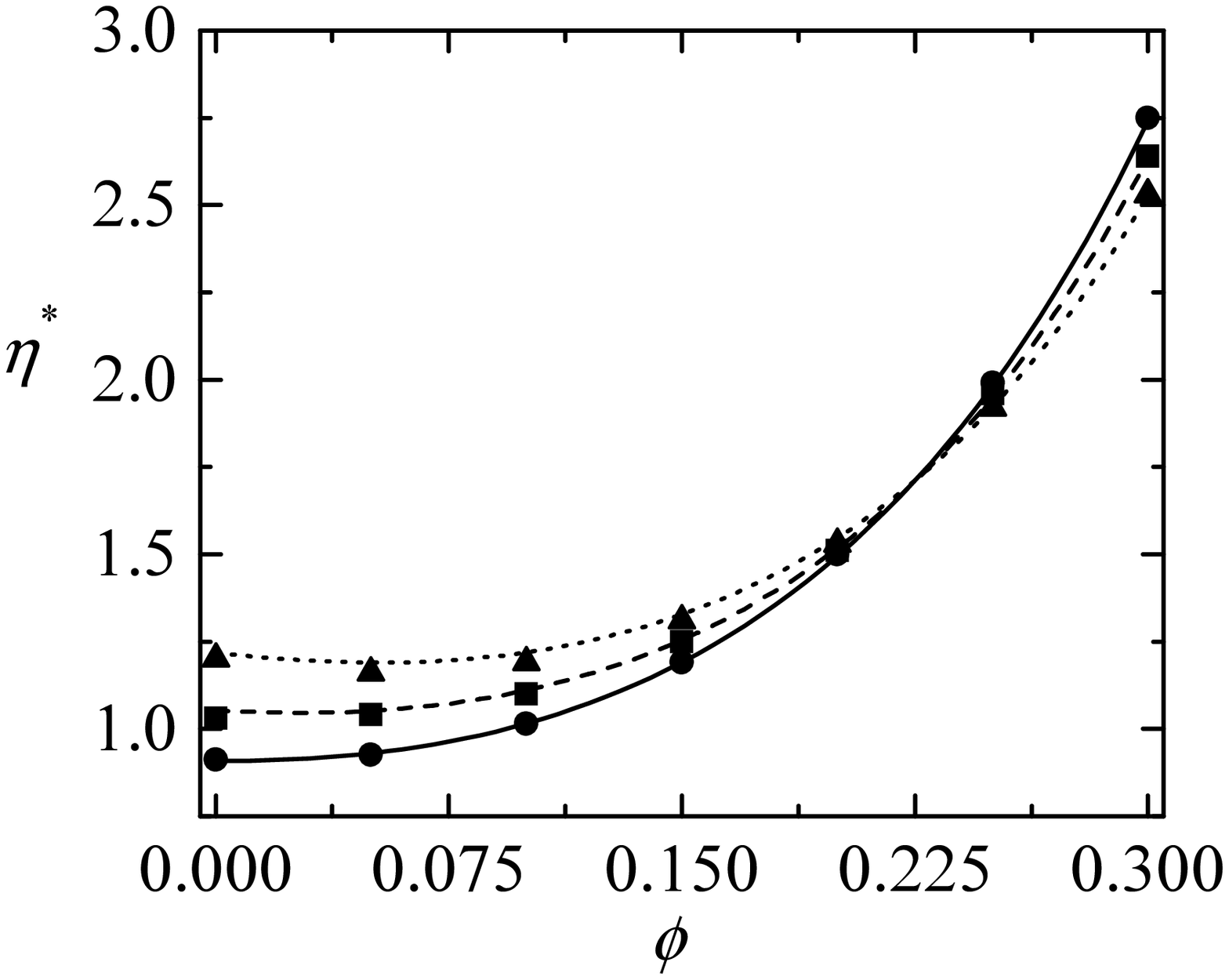,width=6.5cm}\end{center} 
\caption{Plot of the reduced shear viscosity $\eta^*$ as a function of the solid fraction $\phi$ for $\mu=4$, $\omega=\delta=1$ and three different values of the restitution coefficient $\alpha$: $\alpha=0.9$ (solid line and circles), $\alpha=0.8$ (dashed line and squares), and $\alpha=0.7$ (dotted line and triangles). The lines are the theoretical predictions and the symbols refer to the simulations.}
\label{fig11}
\end{figure}

In order to analyze density effects on the shear viscosity, in Figs.\ \ref{fig10} and \ref{fig11} we plot the kinetic part and the total value of this quantity versus the volume packing fraction. The parameters of the mixture are $\mu=4$, $\omega=1$, and $\delta=1$. Three different values of $\alpha$ are studied: $\alpha=0.9$, 0.8, and 0.7. Figure \ref{fig10} shows the dependence of the kinetic part $\eta^{\textin{k}*}=\nu \eta^{\textin{k}}/nT$ on the solid fraction $\phi$, while the total shear viscosity $\eta^*=\nu \eta^/nT$ is plotted in Fig.\ \ref{fig11}. Here, $\nu=\sqrt{\pi}n\sigma_{12}^2v_0$ is an effective collision frequency. The good agreement between theory and simulation indicates that both kinetic and collisional transfer contributions are given accurately by the first Sonine approximation. As in the monocomponent case (cf.\ Fig.\ 2 in Ref.\ \cite{GM03}), the shear viscosity of a granular mixture decreases (increases) as the inelasticity increases if the solid fraction is larger (smaller) than a given threshold value $\phi_0$. The value of $\phi_0$ depends on the parameters of the mixture although it is practically independent of dissipation. For the mixture considered in Fig.\ \ref{fig11}, $\phi_0(\alpha)\simeq 0.22$.  

\begin{figure}[hbt]
\begin{center}\epsfig{file=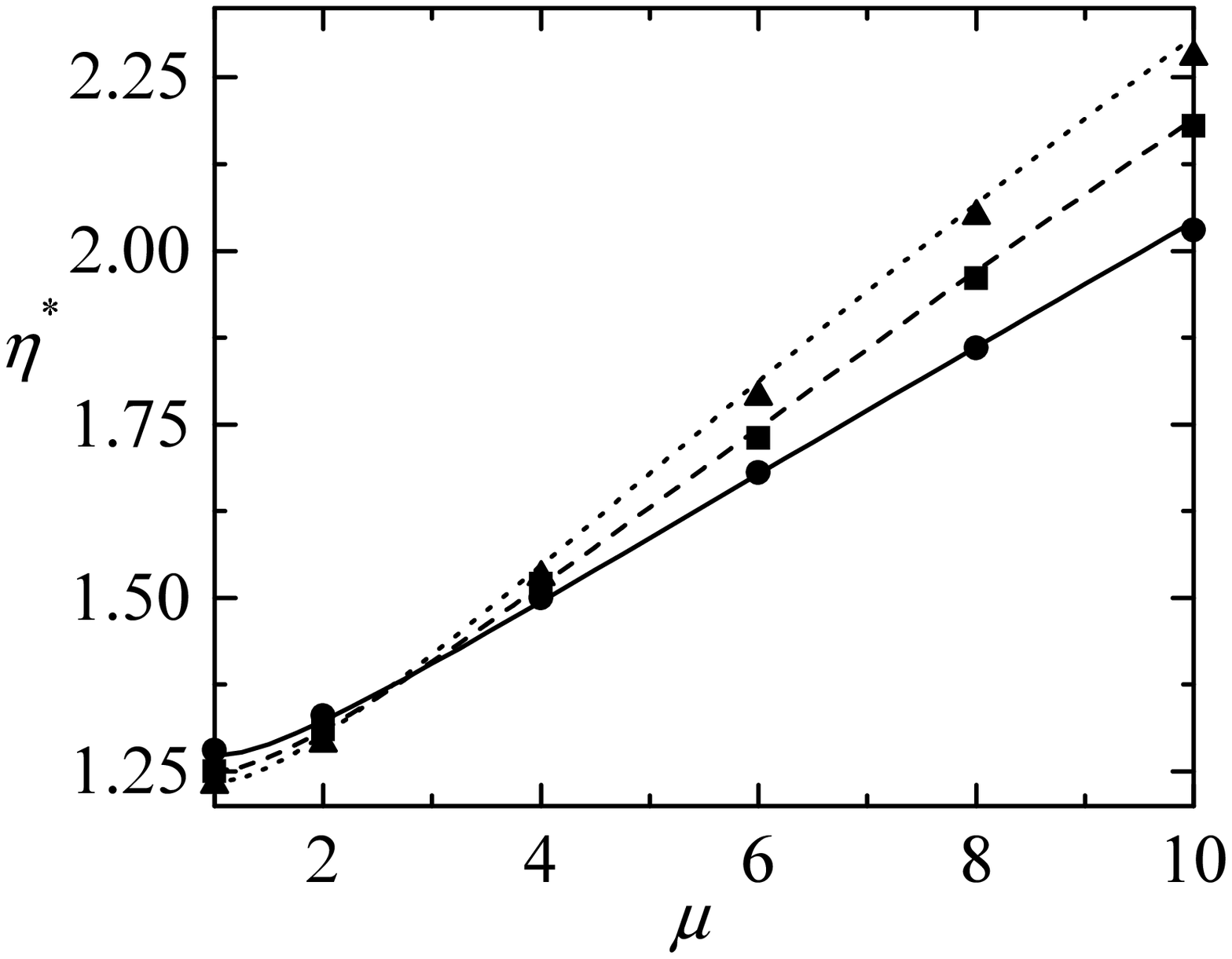,width=6.5cm}\end{center}
\caption{Plot of the reduced shear viscosity $\eta^*$ as a function of the mass ratio $\mu$, for $\omega=\delta=1$, $\phi=0.2$ and three different values of the restitution coefficient $\alpha$: $\alpha=0.9$ (solid line and circles), $\alpha=0.8$ (dashed line and squares), and $\alpha=0.7$ (dotted line and triangles). The lines are the theoretical predictions and the symbols refer to the simulations.}
\label{fig12}
\end{figure}

Finally, we explore in Fig.\ \ref{fig12} the influence of dissipation on the reduced shear viscosity $\eta^*$ for different values of the mass ratio in a mixture at finite density ($\phi=0.2$). As in the low-density case, we see that the influence of dissipation on $\eta^*$ becomes important as the mass disparity increases. At a given value of the mass ratio, $\eta^*$ decreases (increases) with dissipation if the mass ratio is smaller (larger) than a certain threshold value, the value of which seems to be again practically independent of the restitution coefficient. Regarding the comparison between kinetic theory and simulation, we see that the agreement between both approaches is similar to the one previously obtained, although the discrepancies tend to increase as $\alpha$ decreases. 
 
\section{Concluding remarks}
\label{sec5}

In this contribution we have presented an overview of recent results on Monte Carlo simulations of a granular binary mixture. We have considered two homogeneous states: the so-called homogeneous cooling state (HCS) and the stationary state achieved when a gaussian thermostat force is introduced (HSS). It is well known that the results obtained for the velocity distribution functions in both cases are identical when they and velocities are conveniently scaled \cite{MS00,GD99bis,MG02}. Transport properties (such as the shear viscosity coefficient) are related to the response of the granular fluid when small spatial gradients are introduced. In this case, the external forces used as thermostats do not play a neutral role. For instance, Fig.\ \ref{fig6} shows that the discrepancy between the shear viscosity of the driven and undriven cases is quite significant. It must be noticed that the above conclusion only affects to this type of external forcing mechanisms, since driving the system by shaking, vibration, and even the action of a weak external field (such as the gravity field) does not modify the transport coefficients of the granular fluid.

The main motivation of performing Monte Carlo simulations was to verify the range of validity of the approximate analytical predictions obtained from the Enskog equation by using a Sonine polynomial expansion in both the HCS and the HSS. In general, the agreement found is excellent in all the cases considered, even for large disparities of masses, sizes and concentrations. Discrepancies slightly increase as energy dissipation in collisions increases. This agreement is a further testimony to the validity of a hydrodynamic description for granular media beyond the weak dissipation limit. 

Nevertheless, the analysis based on the Enskog equation (\ref{enskog}) is limited in several aspects. A test of the utility of the Enskog equation at high densities is possible using molecular dynamics simulations. Previous comparisons at the level of partial temperatures \cite{DHGD02} and self-diffusion coefficient \cite{LBD02} indicate that the range of densities for which the Enskog theory applies decreases with increasing dissipation. It must be noted that upon describing the HCS and the HSS by the Enskog equations (\ref{enskog}), it has been implicitly assumed the validity of the ``molecular chaos" hypothesis of uncorrelated binary collisions. However, molecular dynamics simulations of hard disks have shown a non-uniform distribution of impact parameters for high enough dissipation \cite{LMM98}. In addition, there exist long range spatial correlations in density and flow fields which can not be understood on the basis of the Enskog equation \cite{NEBO97}. These two effects are associated with the appearance of the so-called cluster instability \cite{G03} for systems sufficiently large. Since we have simulated directly the spatially uniform equation (\ref{enskog}), such an instability is precluded in the simulations.

The main finding of the analysis carried out for binary mixtures is that, in general, the partial temperatures of each species are different \cite{GD99bis,MG02}, in contrast to what was assumed in previous studies \cite{JM89,Z95,AW98,WA99}. This violation of energy equipartition has been subsequently confirmed in experiments of vibrated granular mixtures \cite{WP02,FM02} and in recent molecular dynamics simulations of the HCS \cite{DHGD02}. The Monte Carlo simulations collected in this contribution, and other ones not shown here, demonstrate that the analytical predictions accurately capture the dependence of the temperature ratio on the parameters of the mixture. 

The violation of energy equipartition has been also found in granular mixtures under simple shear flow \cite{MG02a,MG03b,GM03b,CH02}. The simple shear flow state has been extensively studied for molecular fluids as a prototype problem to analyse {\em nonlinear} transport properties. Nevertheless, the nature of this state is quite different in the case of granular fluids. While for elastic fluids the temperature grows monotonically in time due to viscous heating \cite{MS96,MS96b}, a steady state is possible for granular media when the effect of viscosity is exactly compensated by the collisional cooling. In this case, the system reaches a steady state and the temperature achieves a constant value. This state can be analysed using the algorithm described in Sec.\ \ref{sec3} by taking an arbitrary value of the shear rate $a$ and setting $\zeta=0$. The value of the reduced shear rate $a^*=a/\nu$, which is the uniformity parameter that characterizes the steady state, is a function of the normal restitution coefficients $\alpha_{ij}$. The results presented in Refs.\ \cite{MG02a,MG03b,GM03b,CH02} show again that the temperature ratio is clearly different from unity (as may be expected since the system is out of equilibrium) and strongly depends on the restitution coefficients as well as on the parameters of the mixture. The influence of the temperature differences on the rheological properties was analyzed from the Boltzmann kinetic theory by using a Sonine polynomial expansion. The comparison between kinetic theory and Monte Carlo simulations showed an excellent agreement over the range of parameters investigated.

\vspace{1cm}

Partial support from the MCYT (Spain) through Grant No. ESP2003-02859 is acknowledged.

\end{document}